\documentclass[authoryear]{elsarticle}

\usepackage{graphicx}% Include figure files
\usepackage[T1,T2A]{fontenc}
\usepackage[utf8]{inputenc}
\usepackage[english]{babel}
\usepackage{amsmath}
\usepackage{amssymb}
\usepackage{lineno}
\usepackage{tikz}
\usepackage{multirow}
\usepackage{caption}
\usepackage{natbib}
\usepackage{subcaption}
\usepackage{dcolumn}% Align table columns on decimal point
\usepackage{bm}% bold math
\usepackage{pgfplots}
\pgfplotsset{compat=1.15}
\usepackage{mathrsfs}
\usetikzlibrary{shapes.geometric, arrows}

\tikzstyle{startstop} = [rectangle, rounded corners, minimum width=1.8cm, minimum height=1cm,text centered, draw=black, fill=red!30]

\tikzstyle{arrow} = [thin,->,>=stealth]

\begin{document}

\title{Relativistic feedback mechanism in homogeneous electric fields revisited}% Force line breaks with \\

\author{Eduard Kim}
  \affiliation{Moscow Institute of Physics and Technology, Moscow, 117303, Russian Federation;
   Institute for Nuclear Research of RAS, Moscow 117312}
 %\email{kim.e@phystech.edu}
 
 \author{Alexander Sedelnikov}
  \affiliation{Moscow Institute of Physics and Technology, Moscow, 117303, Russian Federation; Lebedev Physical Institute RAS}
%\email{sedelnikov.as@phystech.edu}

\author{Daria Zemlianskaya}
  \affiliation{%
 Moscow Institute of Physics and Technology, Moscow, 117303, Russian Federation;
 Institute for Nuclear Research of RAS, Moscow 117312
}%
%\email{zemlianskay.d@phystech.edu}

\author{Oraz Anuaruly}
  \affiliation{Moscow Institute of Physics and Technology, Moscow, 117303, Russian Federation; Kurchatov Institute RAS;
 Lebedev Physical Institute RAS}
 %\email{orazanuaruly@gmail.com}

\author{Egor Stadnichuk}
\affiliation{Moscow Institute of Physics and Technology, Moscow, 117303, Russian Federation;
 HSE University, Moscow 101000 Russia}
 %\email{yegor.stadnichuk@phystech.edu}


%\affiliation{
%\affiliation{
%\affiliation{%


\date{June 12, 2023}% It is always \today, today,

\begin{abstract}
Recent results link relativistic runaway electron avalanches (RREA) accelerated by the electric field in thunderclouds to high-energy atmospheric phenomena such as the Terrestrial Gamma-Ray Flashes (TGF). Research shows that the mere existence of runaway electron avalanches is not sufficient to generate TGF. In an attempt to settle this issue, a model of a relativistic feedback mechanism was suggested. In this paper, an analytical kinetic revision of the relativistic feedback mechanism is provided. It was shown that positron and gamma feedback mechanisms arise naturally from dynamics equations of RREA initiated independently by positrons and gamma- quanta. Establishing both mechanisms turned out to be enough to evaluate complete relativistic feedback. The electron avalanche multiplication factor is obtained, followed by a study of the minimal conditions of self-sustainable relativistic feedback in homogeneous electric fields and a discussion of the role of this mechanism in TGF and lightning initiation problems.
\end{abstract}

\maketitle

%\tableofcontents

\section{Keypoints}\label{keypoints}
\begin{itemize}

\item Kinetic approach is used to study dynamics of RREA in homogeneous electric fields. 

\item Relativistic feedback mechanism formulation is analytically built in terms of positron- and gamma-based feedback mechanisms.

\item When associated with TGF and lightning initiation, a self-sustainable RREA generation regime in the model under consideration is not achievable under the observed electric fields.

\end{itemize}

\section{\label{intro}Introduction}

One of the major unresolved difficulties in high-energy atmospheric physics is obtaining the necessary conditions for the generation of lightning in thunderclouds.

Many physical models were developed to describe the accumulation of electric charges and corresponding enhancement of electric fields within a thundercloud, as the latest works can be mentioned  [\cite{di2018aerodynamic, iudin2017lightning, dubinova2015prediction}]. Experimental measurements indicate that the absolute value of the electric field is an order of magnitude lower than the value required for conventional electric breakdown in air [\cite{doi:10.1029/JC086iC02p01187, GurevichUFN2001, doi:10.1029/JC076i021p05003}].
As Gurevich showed [\cite{Gurevich1992}], these fields are sufficient for relativistic charged particles to cause runaway relativistic electron avalanches.

Thunderstorm phenomena are not limited only to lightning. According to recent research, thunderstorms are a natural source of gamma radiation; experiments on detecting cosmic showers show that gamma- ray particle flux increases during thunderstorms. Such phenomena as terrestrial gamma-ray flashes (TGF) and thunderstorm ground enhancement (TGE) observed with satellites from space [\cite{Smith1085, doi:10.1029/2009JA015242}] and with detectors on the ground [\cite{chilingarian2012role}] are shown to be the clear, high-energy manifestations coming from thunderstorms. According to the most recent research [\cite{skeie2022temporal}] TGF bursts always occur before or at the same time as the onset of the optical pulse. These observations suggest the importance of high-energy processes for lightning initiation. 

Finally, thunderstorms generate the most powerful natural terrestrial radio bursts in the VHF range [\cite{iudin2015fractal}] known as Narrow Bipolar Events (NBEs), which are thought to be a precursor of lightning, as in the case of TGF. Observations made by Rison [\cite{rison2016observations}] suggested that NBE are produced by volumetrically distributed positive streamers with apparent speeds close to the speed of light. Based on this work, the mechanism of lightning initiation was proposed [\cite{https://doi.org/10.1029/2020JD033191}]. According to this mechanism, runaway electron avalanches trigger giant streamer bursts, the first stage of streamer-leader transition, indicating the role of RREA in lightning initiation. As a result, clarifications on runaway electron generation must be made in order to compare with observations.

The first significant changes in understanding of runaway electron generation were made by Dwyer [\cite{https://doi.org/10.1029/2007JD009248}]. It turned out that Gurevich's runaway electron avalanches, born only from the cosmic shower particles, could not produce enough bremsstrahlung radiation to form the observed fluxes of TGF. No other powerful enough sources of ionization have been observed; therefore, some electron avalanche flux amplification mechanism was required. Dwyer proposed the relativistic feedback mechanism [\cite{Dwyer2003}] which exponentially increases the avalanche generation and changes the evolution of RREA. Recent research, however, suggests that infinite self-generation of RREA by relativistic feedback (RREA burst) is expected to occur only in the presence of a localized, excessively strong, large-scale atmospheric electric field (electric cell) [\cite{refId0}]. "Infinite generations" \text{ on} a TGF lifespan scale $\sim ms$ [\cite{doi:10.1126/science.264.5163.1313}] implies that the vanishing of the electric field due to relativistic discharge is ignored.

In Sec.\ref{independent}, we introduce our approach to studying positron and gamma feedback independently, as components of full-fledged relativistic feedback. Sec.\ref{interference} presents how relativistic feedback might be revived in terms of its components and their mutual interaction. Discussion of results is provided in Sec.\ref{discussion} and conclusions are in Sec.\ref{conclusion}. 

\renewcommand{\tablename}{Table.}
\renewcommand{\figurename}{Fig.}

\section{Positron- and Gamma- relativistic feedback mechanisms}\label{independent}

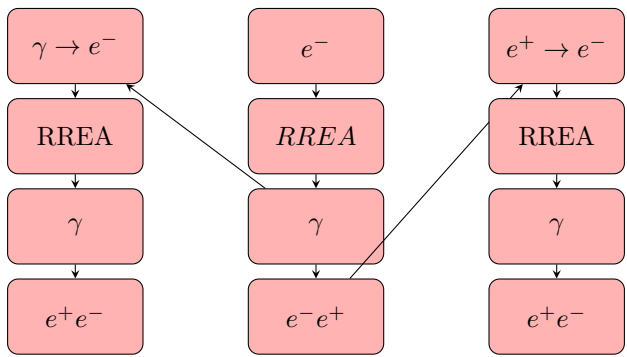
\begin{figure*}
\centering
\begin{tikzpicture}[node distance=1.2cm]
\node (inel) [startstop] {$e^{-}$};
\node (rrea) [startstop, below of=inel] {$RREA$};
\draw [arrow] (inel) -- (rrea);
\node (gamma) [startstop, below of=rrea] {$\gamma$};
\draw [arrow] (rrea) -- (gamma);
\node (pair) [startstop, below of=gamma] {$e^{-}e^{+}$};
\draw [arrow] (gamma) -- (pair);

\node (positron) [startstop, right of=inel, xshift=2cm] {$e^{+}\rightarrow e^{-}$};
\draw [arrow] (pair) -- (positron);
\node (gamma1) [startstop, left of=inel, xshift=-2cm] {$\gamma \rightarrow e^{-}$};
\draw [arrow] (gamma) -- (gamma1);

\node (rrea1) [startstop, below of=positron] {RREA};
\draw [arrow] (positron) -- (rrea1);
\node (gamma2) [startstop, below of=rrea1] {$\gamma$};
\draw [arrow] (rrea1) -- (gamma2);
\node (pair1) [startstop, below of=gamma2] {$e^{+} e^{-}$};
\draw [arrow] (gamma2) -- (pair1);

\node (rrea2) [startstop, below of=gamma1] {RREA};
\draw [arrow] (gamma1) -- (rrea2);
\node (gamma3) [startstop, below of=rrea2] {$\gamma$};
\draw [arrow] (rrea2) -- (gamma3);
\node (pair2) [startstop, below of=gamma3] {$e^{+} e^{-}$};
\draw [arrow] (gamma3) -- (pair2);
\end{tikzpicture}
\caption{Schematic image of positron and gamma relativistic feedbacks producing RREA's second generation in supercritical ($\frac{E}{E_{be}}\geq 1$) electric fields. 
Consideration includes such processes as impact ionization, bremsstrahlung, and electron-positron pair generation. }
\label{proc}
\end{figure*}

In agreement with Fig.\ref{proc}, physical processes are considered in the assumptions presented in Chapter 2 in [\cite{refId0}]. As a result, a set of equations is developed that describes the dynamics of RREA and gamma-quanta in terms of feedback generations.

\begin{equation} \label{eq3.1}
    \begin{cases}
    f_{RREA}^{k+1}=\Hat{\nu}_{e^-}f_{RREA}^{k} + \Hat{\nu}_{\gamma e^-}f_{\gamma}^{k},\\
    f_{\gamma}^{k+1}=\Hat{\nu}_{e^- \gamma}f_{RREA}^{k+1},
    \end{cases}
\end{equation}
where $f$ is a distribution function of RREA or gamma, and operators $\Hat{\nu}$- are integral operators with corresponding eigenvalues $\nu$. Initial conditions of equations (\ref{eq3.1}) are in a form of:

\begin{itemize} \label{initialcond}
    \item $f_{RREA}^{1}=f_0$,
    \item $f_{\gamma}^{0}=0$.
\end{itemize}

According to [\cite{Babich:2020}], short-term disordered development of RREA transfers into a stationary mode of RREA generation. Positron and gamma feedback mechanisms in this section are considered independent mechanisms, thereby integral operators $\Hat{\nu}$ are applicable only to distribution functions $f$ without operator mixing, with only one exception, which will be discussed later. Now equations (\ref{eq3.1}) can be integrated through the parametric space of arguments of distribution functions $f$ and written in the form of 

\begin{equation} \label{eq3.2}
    \begin{cases}
    N_{RREA}^{k+1}=\nu_{e^-}N_{RREA}^{k} + \nu_{\gamma e^-}N_{\gamma}^{k},\\
    N_{\gamma}^{k+1}=\nu_{e^- \gamma}N_{RREA}^{k+1}.
    \end{cases}
\end{equation}

\begin{itemize} \label{initialcond2}
    \item $N_{RREA}^{1}=N_0$,
    \item $N_{\gamma}^{0}=0$.
\end{itemize}

It is critical to discuss the nature of $\nu_{...}$ coefficients before attempting to find the stationary mode of electron avalanche generation. A close examination of the equation (\ref{eq3.1}) reveals that no positron-related processes are taken into account. However, these physics must be factored into $\nu_{e^-}$, which represents a positron feedback mechanism. All gamma generation processes are hidden in $\nu_{e^{-} \gamma}$. Finally, the gamma feedback mechanism is taken into account by $\nu_{\gamma e^{-}}$, which represents the reproduction of RREA by gamma feedback in each generation. 

The first three generations of RREA were derived using equations (\ref{eq3.2}):
\begin{itemize}

    \item First generation
\begin{equation} \label{eq3.3}
    \begin{cases}
    N_{RREA}^{1}=N_0,\\
    N_{\gamma}^{1}=N_0\nu_{e^- \gamma}
    \end{cases}
\end{equation}

    \item Second generation
    \begin{equation}\label{eq3.4}
    \begin{cases}
    N_{RREA}^{2}=N_0(\nu_{e^-}+\nu_{e^- \gamma}\nu_{\gamma e^-}),\\
    N_{\gamma}^{2}=N_0(\nu_{e^-}\nu_{e^- \gamma}+\nu_{\gamma e^-}\nu_{e^- \gamma}^2)
    \end{cases}
\end{equation}

    \item Third generation
    \begin{equation}\label{eq3.5}
    \begin{cases}
    N_{RREA}^{3}=N_0(\nu_{e^-}^2+2\nu_{e^-}\nu_{e^- \gamma}\nu_{\gamma e^-}+\nu_{\gamma e^-}^2\nu_{e^- \gamma}^2),\\
    N_{\gamma}^{3}=N_0(\nu_{e^- \gamma}\nu_{e^-}^2+2\nu_{e^-}\nu_{e^- \gamma}^2\nu_{\gamma e^-}+
    \nu_{\gamma e^-}^2\nu_{e^- \gamma}^3)
    \end{cases}
\end{equation}

\end{itemize}

In a rewritten form, equation (\ref{eq3.5}) takes form 

\begin{equation}\label{eq3.6}
    \begin{cases}
    N_{RREA}^{3}=N_0(\nu_{e^-}+\nu_{\gamma e^-}\nu_{e^- \gamma})^2,\\
    N_{\gamma}^{3}=N_0((\nu_{e^- \gamma}\nu_{e^-}^{2}+2\nu_{\gamma e^-}\nu_{e^- \gamma}^{2}\nu_{e^-})+\nu_{\gamma e^-}^2 \nu_{e^- \gamma}^3)
    \end{cases}
\end{equation}

A close look at equations (\ref{eq3.2})-(\ref{eq3.5}) leads to relations $\frac{N_{RREA}^{4}}{N_{RREA}^{3}}=\frac{N_{RREA}^{3}}{N_{RREA}^{2}}=\frac{N_{RREA}^{2}}{N_{RREA}^{1}}=(\nu_{e^-}+\nu_{e^- \gamma}\nu_{\gamma e^-})$.
Thus, the combined feedback mechanism coefficient for RREA generation is

\begin{equation}\label{eq3.7}
    \Gamma_{in}=(\nu_{e^-}+\nu_{e^- \gamma}\nu_{\gamma e^-})
\end{equation}

In accordance with what is written above, the first term in (\ref{eq3.7}) represents positron feedback, while the second term describes gamma feedback; their explicit forms will be shown in the latter text. To validate the derived formula, the fourth generation of feedback must be compared to the third generation.

\begin{equation}\label{eq3.8}
    N_{RREA}^{4}=N_0(\nu_{e^-}+\nu_{\gamma e^-}\nu_{e^- \gamma})^3
\end{equation}

\begin{equation} \label{eq3.9}
    \Gamma_{in}N_{RREA}^{3}=N_0(\nu_{e^-}+\nu_{\gamma e^-}\nu_{e^- \gamma})^3
\end{equation}

As expected, consistency of equations (\ref{eq3.3}-\ref{eq3.6}) and (\ref{eq3.8}) in not broken.

\subsection{RREA multiplication factors}\label{multiplication}

In addition to Sec.\ref{independent}, defining $\nu$ coefficients will lead to the final formation of a relativistic feedback mechanism as follows from [\cite{https://doi.org/10.1029/2021JD035278, https://doi.org/10.1029/2011JA017160}]. In accordance with [\cite{stadnichuk2022criterion}] positron feedback is represented in the form of

\begin{equation}\label{eq3.10}
\begin{split}
    &\nu_{e^-}=K_{e^-}\left( e^{\frac{L(\lambda_{anih} - \lambda_{RREA})}{\lambda_{anih}\lambda_{RREA}}}- 1 - \frac{L(\lambda_{anih} - \lambda_{RREA})}{\lambda_{anih}\lambda_{RREA}} \right),
\end{split}
\end{equation}

where $K_{e^-}$=$\frac{P_{e^-;e^{+}} P_{e^+} \lambda_{RREA}}{\lambda_2 \lambda_{\gamma} \lambda_{\gamma \rightarrow e^- e^+}} \left(\frac{\lambda_{RREA} \lambda_{anih}}{\lambda_{anih} - \lambda_{RREA}}\right)^2$, with parameter $\lambda_{RREA}$- the length of the exponential rise of RREA, and $\lambda_{anih}$- positron annihilation length. For now, we will treat any occurring $K$ as a normalization constant, which does not describe any dynamics; later in Sec.\ref{interference} it will be properly introduced from kinetic equations. Let $P_{x}$ be the probability of the particle of type $x$ turning around and producing runaway electrons.

\begin{equation} \label{eq3.11}
    \nu_{e^{-}\gamma}=\frac{\lambda_{RREA}}{\lambda_{\gamma}}\left(e^{\frac{z-z_0}{\lambda_{RREA}}}-1 \right),
\end{equation}
where $z_0$- the starting point of RREA and $\lambda_{\gamma}$- mean length at which gamma is produced by runaway electrons.

Because of the exponential law of growth of RREA, the fact that formula (\ref{eq3.11}) describes a source-function of gamma-birth, which is also given in the form of an exponent, and the dependence of the dynamics on the starting point, the process $\gamma \rightarrow e^{-}$ cannot be simply expressed through a multiplication of $\nu_{e^- \gamma}\nu_{ \gamma e^-}$. The only plausible way is to define $\nu_{\gamma e^-}$ as an eigenvalue of the joint ${\widehat{\nu_{e^- \gamma}\nu_{ \gamma e^-}}}$ operator.

The result of operator $\Hat{\nu_{\gamma e^-}}$ acting alongside with $\Hat{\nu_{e^- \gamma}}$ is shown in [\cite{sedelnikov2022the13643}]

\begin{equation}\label{eq3.12}
\begin{split}
    &\widehat{\nu_{e^- \gamma} \nu_{\gamma e}}f_{RREA}=K_{e^- \gamma,\gamma e} \left( e^{\frac{L(\lambda_{x} - \lambda_{RREA})}{\lambda_x\lambda_{RREA}}} -1- \right.\\
    &\left.-\frac{L(\lambda_{x} - \lambda_{RREA})}{\lambda_x\lambda_{RREA}} \right)f_{RREA},
\end{split}
\end{equation}

where $K_{e^- \gamma, \gamma e}=\frac{P_{\gamma} P_{e^-;\gamma} }{\lambda_{\gamma \rightarrow e} \lambda_{\gamma}} \left(\frac{\lambda_{RREA} \lambda_{x}}{\lambda_{x} - \lambda_{RREA}}\right)^2$, $\lambda_{x}$- the characteristic gamma flux attenuation parameter.

\section{Resurgence of the relativistic feedback}\label{interference}

The co-dependence of positron and gamma feedback is essential for the resurgence of relativistic feedback from its components.
For the purpose of reliability of calculations, equations (\ref{eq3.1}) must be taken as a starting point [\cite{https://doi.org/10.1029/2021JD035278}].

The primary electron avalanche starts at $z_0$.
Parameter $\lambda_{\gamma}$- is the length of a runaway electron before the gamma is born, $\lambda_{+}$- is the length of the gamma before the electron-positron pair is born.

Number of produced gamma quanta inside the interval $[z_0,z]$:

\begin{equation} \label{eq3.13}
    f_{\gamma}(z, z_0) = \frac{\lambda_{RREA}}{\lambda_{\gamma}} \cdot \left( e^{\frac{z - z_0}{\lambda_{RREA}}} - 1 \right)
\end{equation}

Number of produced positrons inside the interval $[z_0,z]$:

\begin{equation} \label{eq3.14}
\begin{split}
    &f_{+}(z, z_0) = \int_{z_0}^{z} f_{\gamma}(\zeta, z_0) \frac{d\zeta}{\lambda_{+}} = \frac{\lambda_{RREA}}{\lambda_{\gamma}\lambda_{+}} \Biggl(\lambda_{RREA} e^{\frac{z - z_0}{\lambda_{RREA}}} -\\
    &-\lambda_{RREA} - (z - z_0) \Biggr)
\end{split}
\end{equation}

Number of gamma quanta deployed within a segment $[z_0,z]$:

\begin{equation} \label{eq15}
    f_{\gamma '} (z, z_0) = P_{\gamma} f_{\gamma} (z, z_0)
\end{equation}

Where $\lambda_{\gamma\rightarrow e}$ is the path length of the gamma before the runaway electron is born. Consider that the number of electrons produced by unfolded gamma quanta varies according to the law $\frac{dN}{dz}=\frac{1}{\lambda_{\gamma\rightarrow e}}e^{-\frac{z}{\lambda_x}}$, where $\lambda_x$ is some characteristic length. Then the number of secondary electron avalanches born at coordinate z in thickness $dz$:

\begin{equation}\label{eq16}
    df_2^{gamma}(z, z_0) = dz \cdot \frac{P_{\gamma} \cdot P_{e^-;\gamma}}{\lambda_{\gamma \rightarrow e}} \cdot \int_z^L d\zeta \frac{\partial f_{\gamma}(\zeta, z_0)}{\partial \zeta} e^{\frac{z - \zeta}{\lambda_{x}}}
\end{equation}

Simultaneously, the following is the number of secondary electron avalanches born in the $z$ coordinate in the thickness $dz$:

\begin{equation}\label{eq17}
df_2^{pos}(z, z_0) = dz \cdot \frac{P_{e^-;e^+} P_{e^+}}{\lambda_{2}} \cdot \int_z^L d\zeta \frac{\partial f_+(\zeta, z_0)}{\partial \zeta} e^{-\frac{\zeta - z}{\lambda_{anih}}}
\end{equation}

Since the dynamics of secondary electron avalanches are no different from those of the primary avalanche, the following iterative equation can be written for subsequent generations:

\begin{equation}\label{eq18}
\begin{split}
&f_i(z, 0) = \Hat{L}f_{i-1}(z,0)=\int_0^L (f_2^{pos}(z, \zeta)+f_2^{gamma}(z, \zeta)) \cdot\\
&\cdot \left(\frac{\partial f_{i - 1}^{pos}(\zeta, 0)}{\partial \zeta}+\frac{\partial f_{i - 1}^{gamma}(\zeta, 0)}{\partial \zeta}\right) d\zeta
\end{split}
\end{equation}

Consistently solving this equation leads to the following: there is a stationary mode of RREA generation with the following multiplication factor:

%\begin{widetext}
\begin{equation}\label{eq19}
\begin{split}
    &\Gamma_{r}= \nu_{e^-}+\nu_{e^- \gamma} \nu_{\gamma e}
    + \frac{1}{2}\ast \left( (\nu_{e^-}+\nu_{e^- \gamma} \nu_{\gamma e}) +\sqrt{[(\nu_{e^-}+\nu_{e^- \gamma} \nu_{\gamma e})^2 + 4\nu_{e^-}\ast \nu_{e^- \gamma} \nu_{\gamma e}]} \right)
\end{split}
\end{equation}
%\end{widetext}

\begin{figure*}
\centering
     \begin{subfigure}{1\textwidth}
         \includegraphics[width=\textwidth]{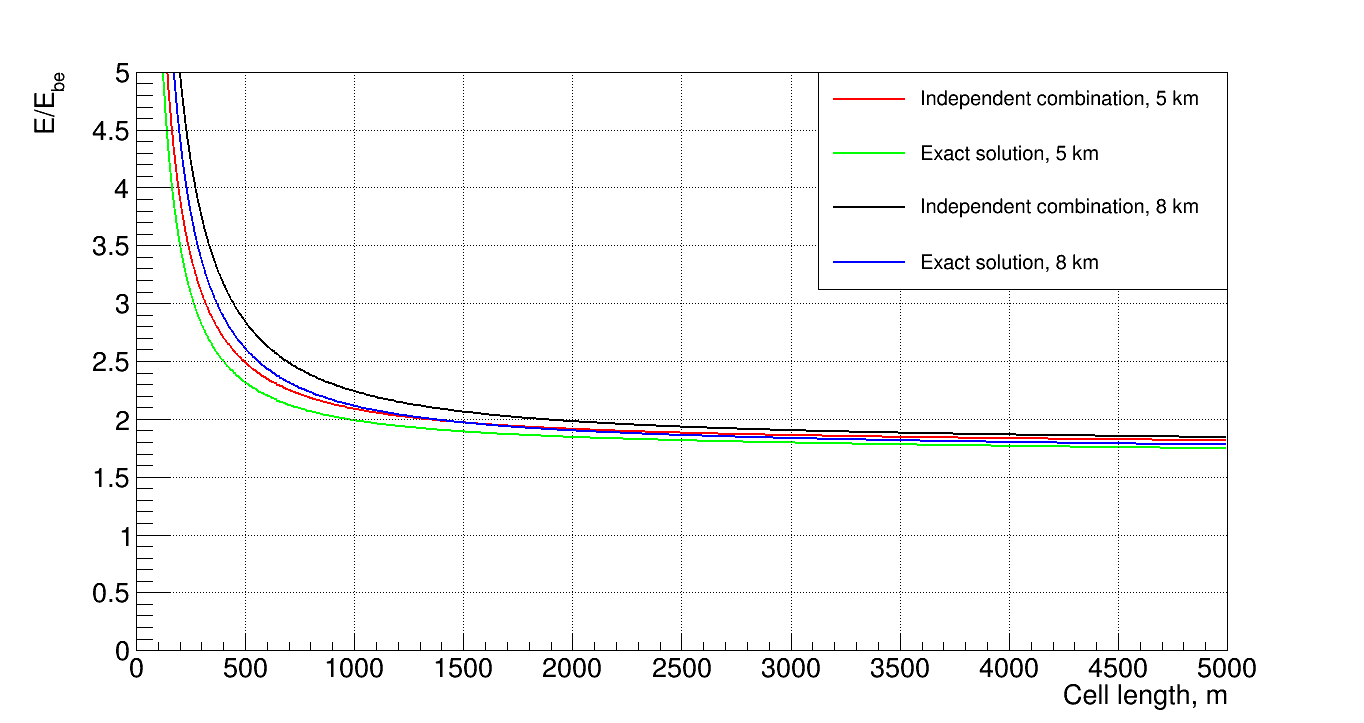}
         \caption{}
         \label{fig2.a}
     \end{subfigure}
     \hfill

     \begin{subfigure}{1\textwidth}
         \includegraphics[width=\textwidth]{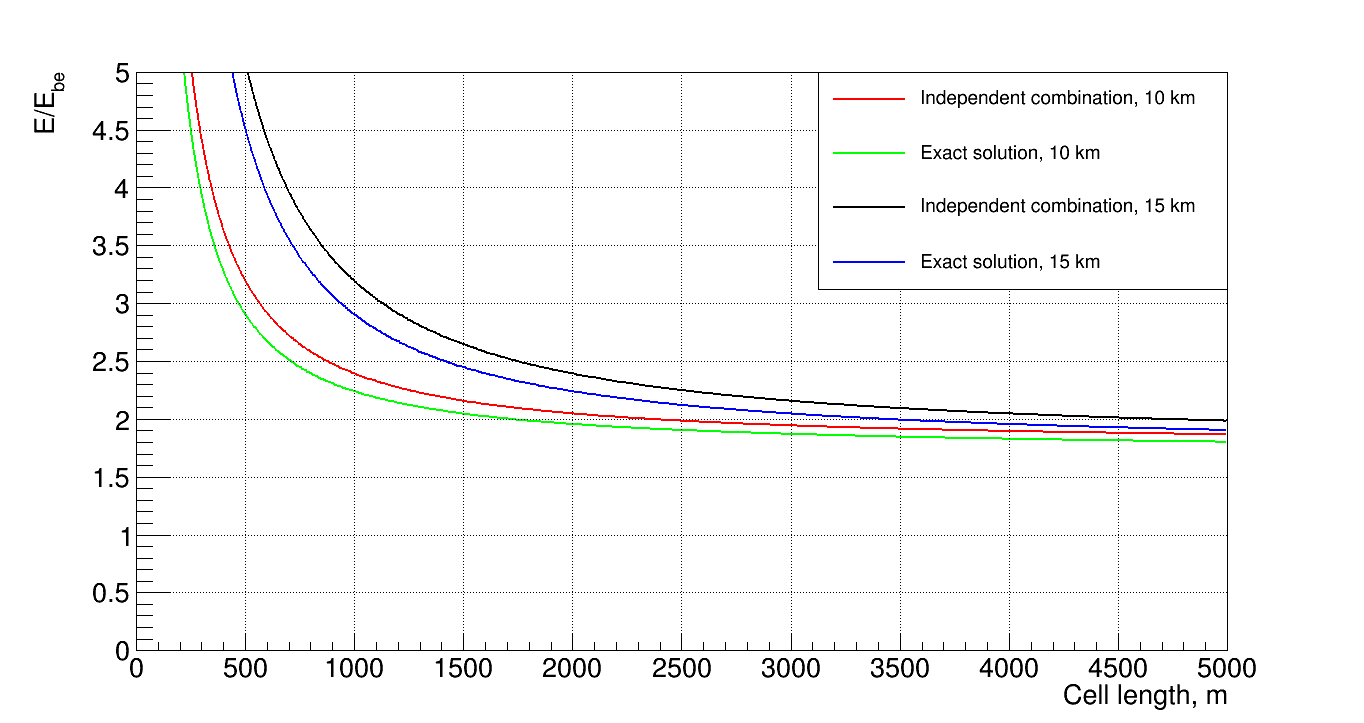}
         \caption{}
         \label{fig2.b}
     \end{subfigure}
     
        \caption{For different altitudes, the minimal conditions for RREA burst ($\Gamma=1$) are calculated in cases of independent combination (positron and gamma feedbacks are taken as independent mechanisms) and the exact solution of the equation (\ref{eq19}), which includes full consideration of feedbacks within their mutual amplification.}
        \label{fig5}
\label{an2}
\end{figure*}

\normalsize
Figure \ref{an2} shows an analysis of the minimal conditions for an infinite RREA burst ($\Gamma=1$) to estimate the difference between equations (\ref{eq3.7}) and (\ref{eq19}). The electric field value is given as a quantity ratio to the break-even field [\cite{article}].

\section{Discussion}\label{discussion}

The equations (\ref{eq3.7}) and (\ref{eq19}) allow the RREA dynamics and RREA burst parameters to be analytically studied in terms of
positron and gamma feedback models. It can be seen from the Fig.\ref{fig2.a}-\ref{fig2.b} that the difference between self-sustaining RREA multiplication regime conditions from equations (\ref{eq3.7}) and (\ref{eq19}) is not significant, and consideration of formula (\ref{eq19}) gives more accurate values of parameters for RREA burst. To have a numerical estimation of the difference considering electric fields is very helpful, and a relation $\frac{|E_{6}-E_{19}|}{E_{19}}=0.0434$, where $E_{6}$ and $E_{19}$ are electric fields from (\ref{eq3.7}), (\ref{eq19}) when
$\Gamma=1$ shows that the difference is only about $4\%$ at an arbitrary altitude. Thereby, it was shown from the kinetic approach that gamma-quanta-induced RREAs and their secondary particles affect relativistic feedback to an insignificant degree. Those results are similar to the ones from [\cite{https://doi.org/10.1029/2011JA017160}] obtained by Monte-Carlo simulations.  

According to [\cite{GurevichUFN2001,https://doi.org/10.1029/2004GL021802}], experimental observations show a correlation between lightning strikes and electric field values close to $E_{be}$. However, the size of a thundercloud region with a homogeneous electric field with $E\geq E_{be}$ was not measured. Cell length could be estimated using the equation (\ref{eq19}) and results from papers such as [\cite{https://doi.org/10.1029/95JD00020, https://doi.org/10.1029/94JD02607}]. Table \ref{table1} was obtained under the assumption that lightning initiation is associated with a RREA burst.

\begin{table}[t]
\begin{tabular}{ |p{1.5cm}|p{1.5cm}|p{1.5cm}|p{1.5cm}||p{1.5cm}|}
 \hline
 \multicolumn{5}{|c|}{Parameters at the moments of lightning discharges with $\Gamma=1$} \\
 \hline
  \multicolumn{3}{|c|}{In situ parameters}&
  \multicolumn{1}{c}{}&
  
  \multicolumn{1}{|c|}{$E/E_{be}=2$}\\
 \hline
 H, km & E, kV/m & $E/E_{be}$ & L, km & $L$, km\\
 \hline
 5.14   & 109   & 0.99 & $4\ast10^{2}$ & 1.16\\
 10     & 59    & 0.939 & $7\ast10^{2}$ & 1.9\\
 11.29  & 48    & 0.89 & $8\ast10^{2}$ & 2.3\\
 7.25   & 75.3  & 0.86 & $5\ast10^{5}$ & 1.4\\
 7.75   & 72.8  & 0.886 & $5\ast10^{5}$ & 1.47\\
 \hline
\end{tabular}
\caption{In situ parameters at the moments of lightning discharges [\cite{https://doi.org/10.1029/95JD00020}] with cell length L derived from equation (\ref{eq19}) under the assumption $\Gamma = 1$. The estimations for L corresponding to the experimentally observed range of $E/E_{be}$ [\cite{https://doi.org/10.1029/2004GL021802}] are provided. Altitude H is given in kilometers, electric field E in $\frac{kV}{m}$ and L in kilometers.}
\label{table1}
\end{table}

In Table \ref{table1} first four rows are filled with parameters ($H$- altitude, $E$- electric field) measured at the moment of lightning strike [\cite{https://doi.org/10.1029/95JD00020}]. In addition, estimation of the upper bound value of $L$ is calculated, given by the experimentally observed range of values of the parameter $\frac{E}{E_{be}}\sim{1 \div 2}$ [\cite{https://doi.org/10.1029/2004GL021802}]. These estimations differ significantly due to the rapid behavior of the minimal conditions for the RREA burst, shown in the figure \ref{an2}. 

To satisfy minimal conditions for RREA burst cell length, it has to be of the order of $10^{2}$ km ($\frac{E}{E_{be}}\sim 1$) within a framework of a relativistic feedback model, or even by a strong overestimation of the values of electric fields ($\frac{E}{E_{be}}=2$) L is of order of $\gtrsim 1$ km. Thus, the obtained results suggest the need for modification of the concept of relativistic feedback for the theory of lightning initiation and TGF. As such modifications, the study of inhomogeneous structures of electric fields in thunderstorms [\cite{https://doi.org/10.1029/2021JD035278}], as well as the possible influence of hydrometeors and their geometry on the generation of runaway electrons [\cite{https://doi.org/10.48550/arxiv.2210.01916}], are considered. The main requirement for such modification should be the mitigation of the RREA burst condition for the observed parameters of the thundercloud.

\section{Conclusion}\label{conclusion}

The aim of this work was to revisit a method of analytical formulation of relativistic feedback from RREA multiplication in the atmosphere. Positron- and gamma-based relativistic feedbacks were proposed to be the building blocks of relativistic feedback. This assumption was proved to be correct as relativistic feedback was revived in the form of $\Gamma=\Gamma(\nu_{e^-}, \nu_{e^- \rightarrow \gamma} \nu_{\gamma \rightarrow e^-})$ 

The influence of RREA's secondary particles (positron and gamma quanta) on its dynamics is expressed not only in their emphasized feedbacks but also in their mutual amplification. Thus, the resulting multiplication factor of resurged relativistic feedback allowed an analysis of the minimal conditions for an RREA burst. Being related to the lightning initiation problem and TGF, relativistic feedback does not provide RREA bursts under the conditions observed in experiments.    

Further research into the influence of feedback mechanisms on RREA dynamics in atmospheric electric field structures will be required to develop a modified feedback mechanism capable of producing RREA bursts at electric field values comparable to those observed in experiments. As well, it is important to study its role in high-energy atmospheric processes, such as TGF, TGE, NBE, etc.  

\section*{Acknowledgement}

The work of Egor Stadnichuk was supported by the Foundation for the Advancement of Theoretical Physics and Mathematics “BASIS”. 

\bibliographystyle{elsarticle-harv}\biboptions{authoryear}
\bibliography{biblio}% Produces the bibliography via BibTeX.

\end{document}